\newcommand{\rem}[1]{}

\documentclass[a4paper,twocolumn]{article}

\usepackage{amsfonts,amssymb,amsmath,amsthm,mathrsfs}
\usepackage{simplewick,hyperref,titlesec,indentfirst,color}
\usepackage[hang,flushmargin]{footmisc} 
\newtheoremstyle{theoremdd}
  {\topsep}
  {\topsep}
  {\itshape}
  {0pt}
  {\bfseries}
  {.}
  { }
  {\thmname{#1}\MakeUppercase{\thmnumber{~\romannumeral #2}}\thmnote{\normalfont{ (#3)}}}
  
\theoremstyle{theoremdd}

\newtheorem{thrm}{Theorem}[section]
\newtheorem{lemma}[thrm]{Lemma}

\newtheorem{cor}[thrm]{Corollary}

\newtheorem{example}[thrm]{Example}
\theoremstyle{definition}

\titleformat{\section}{\center\normalfont\normalsize\bfseries}{\Roman{section}.}{1em}{\MakeUppercase}
\titleformat{\subsection}{\center\normalfont\normalsize\bfseries}{\Alph{subsection}.}{1em}{}

\begin{document}

\title{\bfseries\large NOTES ON WICK'S THEOREM \\
IN MANY-BODY THEORY}
\author{\normalsize Luca Guido Molinari
\\ \small {Dipartimento di Fisica Aldo Pontremoli, Universit\`a degli Studi di Milano,}
\\ \small {and INFN, Sezione di Milano, Via Celoria 16, Milano, Italy}
}
\date{\small 25 October 2017 - revised 13 October 2023}
%
\maketitle
{\bf Abstract}:
In this pedagogical note I present the operator form of Wick's theorem, i.e. a procedure to bring
a product of 1-particle creation and destruction operators to normal order, with respect to some 
reference many-body state. Both the static and the time-ordered cases are presented. For the latter, in particular, 
I provide a simple proof.

\section{\bf Introduction}
For bosons and fermions, destruction and creation operators of a particle 
in a state $|i\rangle$ annihilate the {\em vacuum}: 
$\psi_i|{\rm vac}\rangle $ and $\langle 
{\rm vac}|\psi^\dagger_i=0$. Since observables have null expectation value 
in the vacuum state, it is convenient to construct them with 
creation operators on the left of destruction operators (normal order). 
However, in a many body theory one usually makes reference to the ground state 
$|gs\rangle $ of some non-interacting or effective theory, which is 
filled with particles or quasiparticles. 

Let us suppose that the theory is also 
supplied with a {\em basis of canonical operators} $\alpha_a^-$ and 
$\alpha_a^+ $, $a=1,2,\ldots $ 
\begin{align*}
[\alpha_a^-, \alpha_b^- ]_\mp =0, \quad [\alpha^+_a, \alpha^+_b]_\mp =0
\quad [\alpha_a^-, \alpha_b^+ ]_\mp =\delta_{ab}
\end{align*}
($[O_1,O_2]_\mp =O_1O_2\mp O_2O_1$),
that annihilate the reference state for all $a$:
\begin{align}
\alpha^-_a|gs\rangle =0, \quad \langle gs |\alpha^+_a =0.
\end{align}
Since they are a basis, 1-particle creation or destruction operators $\psi_i$ or $\psi_i^\dagger $, which are hereafter
indifferently denoted as $A_i$, have decomposition
\begin{eqnarray}
A_i=A_i^- + A_i^+
\end{eqnarray}
where the first term is a combination of operators $\alpha_a^-$, and 
the latter is a combination of operators $\alpha_a^+$. 
One term {\sl is not} the adjoint of the other: the labels $-$ and $+$ refer 
to their action on the reference state $|gs\rangle $:
\begin{eqnarray}
A_i^-|gs\rangle =0, \qquad \langle gs |A_i^+ =0 
\end{eqnarray}
Since the operators $\alpha_a^- $ and  $\alpha_a^+ $ are canonical,
by contruction one has
\begin{align}
[A_i^-,A_j^-]_\mp = 0, \quad [A_i^+,A_j^+]_\mp  = 0
\end{align}
while mixed brackets $[A_i^-,A_j^+]_\mp$ are in general non-zero. We  
require them to be c-numbers\footnote{Since $[\alpha_a, \alpha_b^+ ]_\mp =\delta_{ab}$, 
this is certainly true if the operators $A^\pm$ are linear combinations of the $\alpha^\pm_a$.}.  
This is a vital assumption, which makes Wick's theorem 
hold.

In a many particle theory one encounters the problem of expanding 
products of several field operators into normal-ordered expressions 
of the operators $\alpha_a^-$ and $\alpha_a^+ $.
The general problem of bringing products of field operators into a normal 
form was solved in 1950 by Gian Carlo Wick \cite{Wick50} (1909-1992). 
He obtained his theorem while in Berkeley, in the effort to give a clear 
derivation of Feynman's diagrammatic rules of perturbation theory.

\section{Examples} 
\subsection{Non-interacting fermions}
This example is relevant for the perturbation theory with $N$ interacting
fermions. In the zero order description, the two-particle interaction
is turned off and the independent fermions are described by a Hamiltonian 
of the form $H=\sum_a \hbar\omega_a c_a^\dagger c_a$ ($a$ is a label
for one-particle states, ordered so that $\omega_1\le \omega_2\le\ldots $).
The ground state $|F\rangle $ is obtained by filling the 
states $a=1\ldots N$. The operators that annihilate $|F\rangle $ are:
\begin{align}
& \alpha_a^-= \begin{cases} c_a^\dagger  & \text{if $a\le N$,} \\  c_a  & \text{if $a > N$} \end{cases}, 
\quad \alpha_a^- |F\rangle =0 \\
&\alpha_a^+ = \begin{cases} c_a   & \text{if $a\le N$,}\\ 
c_a^\dagger &  \text{if $a > N$}  
\end{cases}, \quad \langle F|\alpha_a^+ =0
\end{align}
$\alpha_a^+$ creates a particle (above the Fermi level)
or creates a hole (by removing a particle in the ``Fermi sea'' $|F\rangle $); $\alpha_a^-$ 
destroys a particle above the Fermi sea, or destroys a hole (by adding a particle in the
Fermi sea). These {\em particle-hole} operators form a CAR. \\
Any destruction or creation operator admits a decomposition in this 
basis into positive and negative parts:
\begin{align*}
\psi_i&=\sum_{a\le N}\langle i|a\rangle  c_a + \sum_{a>N}\langle i|a\rangle c_a  
 = \psi_i^+ +\psi_i^- \\
\psi_i^\dagger&=\sum_{a\le N}\langle a|i\rangle  c_a^\dagger  +
\sum_{a>N}\langle a|i\rangle c_a^\dagger =(\psi_i^\dagger)^-
+(\psi_i^\dagger)^+
\end{align*}
In this example: $(\psi_i^\dagger )^-=(\psi_i^+ )^\dagger $ and $(\psi_i^\dagger )^+ 
=(\psi_i^- )^\dagger $.

\subsection{Bogoliubov transformation}
In this example $|gs\rangle $ is the ground state $|BCS\rangle $ of 
the superconducting state at $T=0$ (Bardeen, Cooper and Schrieffer, 1957). 
It is filled of Cooper pairs of electrons (spin singlets, with zero total momentum), 
\begin{align*}
|BCS\rangle = \prod_{\bf k} (u_k +v_k a_{{\bf k}\uparrow}^\dagger
a_{-{\bf k},\downarrow}^\dagger )|{\rm vac}\rangle
\end{align*}
$u_k $ and $v_k$ are complex amplitudes, with 
$|u_k |^2+|v_k |^2=1$ for the normalization of the state. 
The state is not an eigenstate of the total number operator. 
It is annihilated by the following operators (Bogoliubov and Valatin, 1958):
\begin{align}
\alpha_{\bf k} = u_k a_{{\bf k},\downarrow} + v_k a^\dagger_{-{\bf k},\uparrow}, 
\quad
\beta_{\bf k} = u_k a^\dagger_{{\bf k},\uparrow} - v_k a^\dagger_{-{\bf k},\downarrow} 
\end{align}
\begin{align}
\alpha_{\bf k}|BCS\rangle = 0, \quad \beta_{\bf k}|BCS\rangle =0
\end{align}
Together with their adjoint operators, 
\begin{eqnarray}
\qquad\langle BCS|\alpha_{\bf k}^\dagger = 0, \quad \langle BCS| \beta_{\bf k}^\dagger =0
\end{eqnarray}
they satisfy the CAR rules. They are obtained by a canonical transformation 
that mixes creation and destruction operators of spin-momentum states. 

Inversion gives the operators $a_{{\bf k},\sigma }$ and 
$a^\dagger_{{\bf k},\sigma}$ as sums of a $-$ term (linear combination 
of $\alpha $ and $\beta $) and a $+$ term (linear combination of 
$\alpha^\dagger $ and $\beta^\dagger $ operators). Note that,
although $\langle a_{{\bf k},\sigma}\rangle =0$, BCS-expectation values 
of pairs $aa$ or $a^\dagger a^\dagger $ may be different from zero 
(anomalous correlators).

The variational parameters $u_{\bf k}$ and $v_{\bf k}$ of $|BCS\rangle $ are 
chosen to minimize the ground state energy $\langle H\rangle $. The  
evaluation is simplified by Wick's theorem \cite{Fetter},  which is 
proven in section III.

\subsection{Non-interacting bosons}
For the Bose gas the ground state $|BEC\rangle $ is a Bose-Einstein condensate 
with $N$ particles in the lowest energy state, ${\bf k}=0$, and no particles in higher one-particle 
momentum states, at $T=0$. 
Since $\langle BEC| c_0^\dagger c_0|BEC\rangle =N$, Bogoliubov suggested  
the rescaling $c_0=\sqrt V b$ and $c_0^\dagger = \sqrt V b^*$, 
the other operators being left unchanged. Then
$$ [b,b^* ]=\frac{1}{V}, \quad \langle BEC|b^* b| BEC \rangle = \frac{N}{V}. $$  
The operators $b$ and $b^*$ may be treated as 
c-numbers in the thermodynamic limit \cite{Lieb04}, 
with $|b|^2=N/V$. For the field operators one has the 
decomposition into a condensate term, and an excitation field operator:
\begin{align}
&\psi ({\bf x})= \sum_{\bf k}\langle {\bf x}|{\bf k}\rangle c_{\bf k}=  b  + \phi ({\bf x}), \\
&\psi^\dagger({\bf x})= \sum_{\bf k}\langle {\bf x}|{\bf k}\rangle^* c_{\bf k}^\dagger=  b^*  + \phi^\dagger ({\bf x})
\end{align}
where $\phi ({\bf x}) |BEC\rangle =0$ and $\langle BEC|\phi^\dagger ({\bf x}) =0$.

\section{\bf Normal ordering and contractions}
A product of operators $A_i^\pm$ is
{\sl normally ordered } if all factors $A_i^-$ are at the right of the 
factors $A_j^+$:
\begin{eqnarray}
A_1^+\cdots  A_k^+ A_{k+1}^- \cdots A_n^-
\end{eqnarray}
In particular, a product of operators of the same type, $A_1^+\cdots A_k^+$
or $A_1^-\cdots A_\ell^-$, is normally ordered.
The very usefulness of the definition is the obvious property that
the expectation value on $|gs\rangle $ of a normally ordered operator is 
always zero:
\begin{eqnarray}
\langle gs|A_1^+ \cdots A_n^-|gs\rangle =0
\end{eqnarray}
It is clear that any product of operators $A_1A_2\ldots A_n$ can
be written as a sum of normally ordered terms. One first writes every
factor as $A_i^++A_i^-$ and gets $2^n$ terms. In each term, the components
$A_i^-$ are brought to the right by successive commutations (bosons) or
anticommutations (fermions). After much boring work, the desired expression
will be obtained. 
Wick's theorem is an efficient answer to this precise problem: to write a 
product $A_1\ldots A_n$ as a sum of normally ordered terms. The theorem is
an extremely useful operator identity, with important corollaries. To state 
and prove it,  we need some technical tools.

The {\bf normal ordering operator} brings a generic
product into a normal form. If the product contains $k$ factors $A_i^+$ mixed
with $n-k$ factors $A_i^-$, it is\footnote{Another frequently used notation for normal ordering
is $:A_1^\pm \cdots A_n^\pm :$.}: 
\begin{align}
{\sf N}[A_1^\pm \cdots A_n^\pm ]= (\pm 1)^P A_{i_1}^+\cdots A_{i_k}^+\cdots 
A_{i_n}^-  \label{normalordering}
\end{align}
For bosons $(+1)^P=1$; for fermions  $(-1)^P$ is the parity of the permutation 
that brings the sequence $1\ldots n$ to the sequence $i_1\ldots i_n$. \\
It may appear that normal 
ordering is not unique, since within
$+$ operators or $-$ operators one can choose different orderings.
However the different expressions are actually the same operator,
because $A^+$ operators commute or anticommute exactly among themselves, 
and the same is for $A^-$ operators. For example ${\sf N}[A_1^+A_2^+]$  can be 
written either $A_1^+ A_2^+$ or with factors exchanged: $\pm A_2^+A_1^+$, the two outputs coincide.

The action of $N$-ordering is extended by linearity from
products of components $A_i^\pm $  to products of operators $A_i$. 
For example:
\begin{align*}
&{\sf N}[A_1A_2]= {\sf N}[(A_1^+ +A_1^- )(A_2^+ +A_2^- )]\\
&= {\sf N}[A_1^+A_2^+ ]+ {\sf N}[A_1^+ A_2^- ]+{\sf N}[A_1^- A_2^- ]+{\sf N}[A_1^- A_2^+ ] \\
&= A_1^+ A_2^+ + A_1^+ A_2^- + A_1^- A_2^- \pm  A_2^+ A_1^-
\end{align*} 
The following property follows from \eqref{normalordering}:
\begin{align}
{\sf N}[A_1\cdots A_n] = (\pm 1)^P\, {\sf N}[A_{i_1}\cdots A_{i_n}]
\end{align}

A product $A_1\ldots A_n$ can be written as a sum of normally ordered terms.
For two operators the process is straightforward:
\begin{align*}\label{eq:NA1A2}
A_1A_2 =& (A_1^+ +A_1^-)(A_2^++A_2^-)\nonumber \\
=& A_1^+A_2^+ + A_1^+A_2^- + A_1^-A_2^- + A_1^-A_2^+\nonumber \\
=& {\sf N}[A_1A_2]+ [A_1^-,A_2^+]_\mp
\end{align*}
The last term $[A_1^-,A_2^+]_\mp  $ is  a c-number 
that defines the {\bf contraction}, denoted by a bracket, of two operators:
\begin{equation}
\contraction{}{A_1}{}{A_2}{A_1}{A_2}=[A_1^-,A_2^+]_\mp
\end{equation}
\begin{align}
 A_1A_2 = {\sf N}[A_1A_2]  + \contraction{}{A_1}{}{A_2}{A_1}{A_2} 
\end{align}
Since the $gs-$expectation value of a normal ordered operator
is zero, it follows that
\begin{equation}
\contraction{}{A_1}{}{A_2}
{A_1}{A_2} = \langle gs|\contraction{}{A_1}{}{A_2}
{A_1}{A_2}|gs \rangle =\langle gs| A_1A_2| gs\rangle
\end{equation}
The following definition extends the contraction of two operators to the case
where there is a product of $n$ operators in between:
\begin{eqnarray}
\contraction{}{A}{(A_1\cdots A_n)}{A'}A{(A_1\cdots A_n)}{A^\prime}
 = (\pm 1)^n \contraction{}{A}{}{A^\prime}{A}{A^\prime} (A_{1}\cdots A_{n}) 
\end{eqnarray}
\section {\bf Static Wick's theorem}
We begin by proving three Lemmas; each one is a generalization of the former. 
In the first one, a single $A^-$ operator is at the left of $A^+$ operators, and normal ordering is achieved by 
bringing it to the right of them by repeated (anti)commutations.
\begin{lemma}\label{Lemma}
\begin{align}
 &A_0^- A_1^+\cdots A_n^+  \\
 & ={\sf N}[A_0^-A_1^+\cdots A_n^+]+ \sum_{i=1}^n {\sf N}[
\contraction{}{A}{_0\cdots }{A} {A_0\cdots A_i}\cdots A_n^+] \nonumber
\end{align}
\begin{proof}
\begin{align*}
& A_0^- A_1^+\cdots A_n^+\\
& =([A_0^-,A_1^+]_\mp)A_2^+\cdots A_n^+ \pm A_1^+A_0^-A_2^+\cdots A_n^+ \\
& =\contraction{}{A_0}{}{A_1}{A_0}{A_1}
A_2^+\cdots A_n^+ \pm A_1^+([A_0^-, A_2^+]_\mp)
A_3^+\cdots A_n^+\\
&\quad + A_1^+ A_2^+ A_0^-A_3^+\cdots A_n^+  \\
& = \contraction{}{A_0}{}{A_1}{A_0}{A_1} A_2^+\cdots A_n^+ + 
\contraction{}{A_0}{A_1^+}{A}{A_0A_1^+A_2}
A_3^+ \cdots A_n^+\\
&\quad + A_1^+ A_2^+ A_0^-A_3^+ \cdots A_n^+ =\ldots \\
& = \sum_{i=1}^n \contraction{}{A}{_0A_1^+\cdots}{A} {A_0 A_1^+\cdots A_i} \cdots A_n^+ +
(\pm 1)^n A_1^+\cdots A_n^+ A_0^- 
\end{align*}
The last term is precisely $ {\sf N}[A_0^-A_1^+\cdots A_n^+]$.
\end{proof}
\end{lemma}
\begin{lemma}\label{Lemma 2}
\begin{align}
& A_0^- {\sf N}[A_1\cdots A_n] \label{L2}\\
& = {\sf N}[A_0^- A_1\cdots A_n] + \sum_{i=1}^n  {\sf N}[
\contraction{}{A}{_0\cdots }{A} A_0\cdots A_i
 \cdots A_n].
\nonumber
\end{align}
\begin{proof}
The proof is by induction. Eq.\eqref{L2} holds for $n=1$. If, by hypothesis, it holds for 
$n$ operators, it is now proven for $n+1$ operators (we write $1^{\pm}$ in place of $A_1^\pm $): 
\begin{align*}
&0^-{\sf N}[1\cdots (n+1)] \\
&=0^-1^+{\sf N}[2\cdots (n+1)]+(\pm 1)^n
  0^-{\sf N}[2\cdots (n+1)1^-]\\
&=\contraction{}{0}{}{1}{0}{1} \,
{\sf N}[2 \,\cdots ]\pm 1^+0^-{\sf N}[2\,\cdots]+(\pm 1)^n 0^- 
{\sf N}[2,\cdots] 1^-
\end{align*}
The hypothesis of induction is now used in the second and third terms:
\begin{align*}
= \, & {\sf N}[\contraction{}{0}{}{1}{0}{1} \,
 2\cdots (n+1)]\\
 & \pm 1^+   \big \{  {\sf N}[0^-2\cdots (n+1)] 
+ \sum_{k\ge 2}  {\sf N}[
\contraction{}{0}{\cdots}{k}{0}{\cdots}{k}
\cdots (n+1)] \big \} \\
& (\pm 1)^n \big \{ {\sf N}[0^- 2\cdots (n+1)] 
+  \sum_{k\ge 2} {\sf N}[
\contraction{}{0}{\cdots}{k}{0}{\cdots}{k}
\cdots (n+1)]\big \}  1^- \\
=\, &{\sf N}[\contraction{}{0}{}{1}{0}{1}
2\cdots (n+1)] \pm {\sf N}[1^+0^-2\cdots (n+1)]\\
&+\sum_{k\ge 2}{\sf N}[
\contraction{}{0}{1^+2\cdots}{k}{0}{1^+2\cdots}{k}
\cdots (n+1)]\\
&+ (\pm 1)^n{\sf N}[0^-  2 \cdots (n+1)1^-]\\
&+(\pm 1)^n \sum_{k=2..n+1} {\sf N}[
\contraction{}{0}{2\cdots}{k}{0}{2\cdots}{k}
 \cdots  (n+1)1^-]\\
=\, &{\sf N}[\contraction{}{0}{}{1}{0}{1}
2\cdots (n+1)] + {\sf N}[0^-1^+2\cdots (n+1)]\\
&+\sum_{k\ge 2}{\sf N}[
\contraction{}{0}{1^+\cdots}{k}{0}{1^+\cdots}{k}
\cdots (n+1)]\\
&+{\sf N}[0^- 1^- 2 \cdots (n+1)]
+ \sum_{k\ge 2} {\sf N}[
\contraction{}{0}{1^-2\cdots}{k}{0}{1^-2\cdots}{k}
 \cdots  (n+1)]\\
=\, &{\sf N}[0^-1 \cdots (n+1)]+\sum_{k\ge 1}{\sf N}[
\contraction{}{0}{1\cdots}{k}{0}{1\cdots}{k}
\cdots  (n+1)].
\end{align*}
The last line is \eqref{L2} with $n+1$ operators.
\end{proof}
\end{lemma}

\begin{lemma}\label{Lemma 3}
\begin{align}
A_0 {\sf N}[A_1\cdots A_n] 
= {\sf N}[A_0\cdots A_n] + \sum_{i=1}^n 
{\sf N}[
\contraction{}{A_0}{\cdots}{A} A_0\cdots A_i
 \cdots A_n].  \nonumber
\end{align}
\begin{proof}
This is achieved by adding $A^+_0{\sf N}[A_1\cdots A_n]={\sf N} [A_0^+ A_1\cdots A_n]$ to both sides of Lemma \ref{Lemma 2}.
\end{proof}
\end{lemma}

Wick's theorem gives the practical rule to express a
product of creation and destruction operators as a sum of
normally ordered terms. It is an operator identity.\\
Each contraction, being a c-number,
reduces by two the operator content.
\begin{thrm}[Static Wick's Theorem]
\begin{align}
& A_1 A_2 \cdots A_n = {\sf N}[A_1 \cdots A_n]  \label{WT} \\
&+\sum_{(ij)} {\sf N}[A_1\cdots \contraction{}{A_i}{ \cdots }{A_j} A_i \cdots A_j
\cdots  A_n]\nonumber\\
& +\sum_{(ij)(kl)}  {\sf N}[A_1\cdots 
\contraction{}{A}{_i\cdots A_k \cdots}{A_j}
\contraction[1.5ex]{A_i\cdots}{A}{_k\cdots A_j\cdots }{A_l}
A_i\cdots A_k \cdots A_j \cdots A_l \cdots A_n]  \nonumber \\
& + \ldots             \nonumber
\end{align}
The first sum runs on single contractions of pairs, 
the second sum runs on
double contractions, and so on.\\ 
If $n$ is even, the last sum contains 
terms which are products of contractions (c-numbers). If $n$ is odd, the last sum
has terms with single unpaired operators (see examples).
\begin{proof}
The theorem is proven by induction. For $n=2$ it is true. Next, 
suppose that the statement is true for a product of creation/destruction 
operators $A_1\cdots A_n$: it is shown that 
it is true for a product $A_0A_1\cdots A_n$. \\
By hypothesis of induction for $n$ operators:
\begin{equation}
A_1\cdots A_n = \sum_{k=0}^{ \lfloor n/2\rfloor} N_{n,k}
\end{equation}
where $N_{n,k}$ is the sum of normally ordered products of $n$ operators with $k$ 
contractions\footnote{For example,  $N_{n,2} = \sum_{(pq)(rs)}  {\sf N}[A_1... 
\contraction{}{A}{_p... A_r ...}{A_q}
\contraction[1.5ex]{A_p...}{A}{_r... A_q... }{A_s}
A_p... A_r ... A_q ... A_s ... A_n]
$.}.
By Lemma \ref{Lemma 3}: 
\begin{equation}
A_0 N_{n,k} = N[A_0 N_{n,k}] + N[\contraction{}{A}{_0{}}{N} A_0 N_{n,k}],
\end{equation}
where $\contraction{}{A}{_0{}}{N} A_0 N_{n,k}$ means the sum of all contractions of $A_0$ with
unpaired operators $A_i$ contained in $N_{n,k}$.\\
The following relation takes place: 
\begin{equation} N[\contraction{}{A}{_0{}}{N} A_0 N_{n,k}] + N[A_0 N_{n,k+1}]=N_{n+1,k+1}.
\end{equation}
Using the induction hypothesis and the last two identities we find,
\begin{align*}
& A_0A_1\cdots A_n = A_0 N_{n,0}+ A_0 N_{n,1}+ A_0 N_{n,2}+\dots\\
& 
 = {\sf N}[A_0 N_{n,0}] +{\sf N} [\contraction{}{A}{_0}{N}A_0 N_{n,0} ]+
{\sf N}[A_0 N_{n,1}] + {\sf N}[\contraction{}{A}{_0}{N}A_0 N_{n,1}] \\
&\; 
\quad +{\sf N}[A_0 N_{n,2}] + {\sf N}[\contraction{}{A}{_0}{N}A_0 N_{n,2} ]+ \dots \\
& = N_{n+1,0} + N_{n+1,1}+\dots
\end{align*}
which expresses Wick's theorem for $n+1$ operators.
\end{proof}
\end{thrm}

\begin{example}
\begin{align}
 A_1A_2A_3 &= {\sf N}[123] + {\sf N}[\contraction{}{1}{}{2}123] + {\sf N}[\contraction{}{1}{2}{3}123]+{\sf N}[1\contraction{}{2}{}{3}23]\nonumber \\ 
&= {\sf N}[123]+\langle 12\rangle A_3 \pm \langle 13\rangle A_2
+ \langle 23\rangle A_1 \label{AAA}
\end{align}
\begin{align}
& A_1A_2A_3A_4 \nonumber \\
& = {\sf N}[1234] + {\sf N}[\contraction{}{1}{}{2}1234] + {\sf N}[\contraction{}{1}{2}{3}1234] + {\sf N}[\contraction{}{1}{23}{4}1234]
\nonumber \\
&\quad + {\sf N}[1\contraction{}{2}{}{3}234] + {\sf N}[1\contraction{}{2}{3}{4}234] + {\sf N}[12\contraction{}{3}{}{4}34]
\nonumber \\
&\quad +{\sf N}[\contraction{}{1}{}{2}12\contraction{}{3}{}{4}34]+{\sf N}[\contraction{}{1}{2}{3}\contraction[1.5ex]{1}{2}{3}{4}1234]+{\sf N}[\contraction{1}{2}{}{3} \contraction[1.5ex]{}{1}{23}{4}
1234]
\nonumber \\
&={\sf N}[1234]+\langle 12\rangle {\sf N}[34] \pm \langle 13\rangle {\sf N}[24]+
\langle 14\rangle {\sf N}[23]
\nonumber\\
&\quad + \langle 23\rangle {\sf N}[14] \pm \langle 24\rangle {\sf N}[13]
\pm \langle 34\rangle {\sf N}[12]
\nonumber \\
&\quad +\langle 12\rangle \langle 34\rangle 
\pm \langle 13\rangle \langle 24\rangle +\langle 14\rangle\langle 23\rangle
\label{AAAA}
\end{align}
\end{example}

An important consequence of Wick's operator identity is a rule 
for the expectation value of the product of an even number of destruction 
and creation operators:
\begin{cor}
\begin{align}
&\langle gs|A_1 \cdots A_{2n} |gs\rangle \\
&= \sum (\pm 1)^P 
 \langle gs|A_{i_1} A_{j_1}|gs \rangle\cdots \langle gs|A_{i_n} A_{j_n}|gs
\rangle \nonumber 
\end{align}
The sum is over all partitions of $1,\ldots ,2n$ into
pairs $\{(i_1,j_1)\ldots (i_n,j_n)\}$ with $i_\#<j_\#$. $P$ is the permutation that takes $1,\dots , 2n$ to
the sequence $i_1,j_1,\dots ,i_n,j_n$.
\end{cor}

\begin{example}
\begin{align} 
&\langle gs|123|gs\rangle =0\\
&\langle gs|1234|gs\rangle = \langle 12\rangle \langle 34\rangle 
\pm \langle 13\rangle \langle 24\rangle +\langle 14\rangle\langle 23\rangle
\end{align}
\end{example}
This is a general rule: {\em two-point correlators determine all n-point
correlators}.\\

In thermal theory there is no distinguished state to define a normal ordering, 
and thus no Wick's theorem in the form of an operator identity.
Nevertheless, one can prove a thermal analogue of the corollary. 
For {\sl non-interacting} particles the thermal average of a product of 
one-particle creation and destruction operators is the sum of all possible
thermal contractions of pairs. The thermal contraction 
of two operators is the thermal average of their product \cite{Fetter,Gaudin}.

\section{Wick's theorem with time-ordering}
An important variant of Wick's theorem deals with the 
normal-ordering of a {\sl time-ordered product}. A necessary condition 
is that the time evolution of the operators $\alpha_a^- $ and $\alpha_a^+ $ is
a multiplication by some time-dependent phase factor (c-number).
Then, the discussion on normal ordering and contraction of operators $A_i(t_i)$
remains unaltered. 
\subsection{T-contractions}
Let us begin with two 
operators, and apply Wick's theorem: 
\begin{align}
&{\sf T}A_1(t_1)A_2(t_2)\nonumber \\
&=\theta(t_1-t_2)\{{\sf N}[A_1(t_1)A_2(t_2)]+
\contraction{}{A}{_1(t_1)}{A}A_1(t_1)A_2(t_2) \}\nonumber \\
&\quad\pm \theta(t_2-t_1)\{{\sf N}[A_2(t_2)A_1(t_1)]+\contraction{}{A}{_2(t_2)}{A}A_2(t_2)A_1(t_1)\}
\nonumber \\
&=\theta(t_1-t_2){\sf N}[A_1(t_1)A_2(t_2)]+ \theta(t_2-t_1){\sf N}[A_1(t_1)A_2(t_2)]
\nonumber \\
&\quad +\theta(t_1-t_2) \contraction{}{A}{_1(t_1)}{A}A_1(t_1)A_2(t_2) \pm \theta(t_2-t_1)
\contraction{}{A}{_2(t_2)}{A}A_1(t_1)A_2(t_2) \nonumber \\
&= {\sf N}[A_1(t_1)A_2(t_2)]+\overbrace{A_1(t_1)A_2(t_2)} \label{YYY}
\end{align}
The last term is a c-number and defines the time-ordered contraction ({\sf T}-contraction). It is
\begin{eqnarray}
\overbrace{A_1(t_1)A_2(t_2)} = \langle gs|{\sf T} A_1(t_1)A_2(t_2)|gs\rangle 
\end{eqnarray}
The {\sf T}-contraction of two operators with a product of $n$ operators in between 
inherits the property of static contractions
\begin{eqnarray}
\overbrace{A_1(t_1)(\cdots )A_2(t_2)}=(\pm 1)^n
\overbrace{A_1(t_1)A_2(t_2)}(\cdots )
\end{eqnarray}
{\sf T}-contractions have a new property, not shared by a static contraction:
\begin{eqnarray}
\overbrace{A_1(t_1)A_2(t_2)} = \pm \overbrace{A_2(t_2) A_1(t_1)}
\end{eqnarray}
For field operators we have the explicit expressions:
\begin{align}
&\overbrace{\psi (1)\psi^\dagger (2)}=\langle gs| {\sf T}\psi(1)\psi^\dagger (2)|gs \rangle =
iG^0(1,2)\\
&\overbrace{\psi(1)\psi(2)}=\langle gs| {\sf T}\psi(1)\psi(2)|gs \rangle = iF^0(1,2),\\ 
&\overbrace{\psi^\dagger (1)\psi^\dagger (2)}=\langle gs| {\sf T}\psi^\dagger(1) 
\psi^\dagger (2)|gs \rangle = i F^{\dagger 0}(1,2)
\end{align}
If $|gs\rangle $ has a definite number of particles, the
{\sl anomalous} correlators $F^0$ and $F^{\dagger 0}$ are equal to zero. They
are non-zero in the BCS theory. 

\subsection{Wick's theorem for time-ordered products}
For the time-ordered product of several operators, Wick's theorem 
retains the same structure as in (\ref{WT}), with {\sf T}-contractions replacing 
ordinary ones. The statement is:

\begin{thrm}[Wick's theorem with time-ordering]
\begin{align}
& {\sf T}[A_1(t_1) \cdots A_n (t_n)]= {\sf N}[A_1(t_1) \cdots A_n(t_n)]  \label{WTTO} \\
&+\sum_{(ij)} {\sf N}[ A_1(t_1)\cdots \overbrace{A_i(t_i) \cdots A_j(t_j)} \cdots  A_n(t_n)]\nonumber\\
& +\sum  {\sf N}[\;\cdots \text{\rm double {\sf T}-contractions} \cdots \;]  + \ldots\nonumber
\end{align} 
\begin{proof} The proof is by induction. For $n=2$ it is \eqref{YYY}. Now suppose that it is true for $n$
(we omit the specification of time):
\begin{align*}
 {\sf T}[A_1 \cdots A_n]= N_{n,0}+ N_{n,1}+N_{n,2}+ \ldots 
 \end{align*}
where $N_{n,k}$ is the term in \eqref{WTTO} with $k$ time-ordered contractions.  
Consider the ${\sf T}$ product of $n+1$ operators at different times and let $A_\ell $ in the product be the operator at largest time:
\begin{align}
 {\sf T}[A_0 A_1 \cdots A_n]= (\pm 1)^\ell A_\ell  {\sf T}[A_0 A_1 \cdots A_n]_\ell 
 \end{align}
 $[A_0\ldots A_n]_\ell $ means that $A_\ell $ is absent. The ${\sf T}$ product now acts on $n$ operators and the hypothesis applies:
$$ (\pm 1)^\ell A_\ell  {\sf T}[A_0 A_1 \cdots A_n]_\ell = (\pm 1)^\ell A_\ell (N_{n,0}+ N_{n,1}+...)$$ 
Now use Lemma IV.3 for each $k$:
$$  A_\ell N_{n,k} =  {\sf N}[A_\ell N_{n,k}] +  {\sf N} [\contraction{}{A}{_\ell}{N}A_\ell N_{n,k} ] $$
In the second term, since $t_\ell$ is maximal, the contraction of $A_\ell$ with the available operators in $N_{n,k}$ coincides with a time ordered contraction, 
$\contraction{}{A}{_\ell}{A}A_\ell A_j  = \langle gs |{\sf T} A_\ell A_j |gs\rangle $. \\
In the following sum, the operator $A_\ell $ is moved to its place with $\ell$ exchanges:
\begin{align}
 (\pm 1)^\ell {\sf N}[A_\ell N_{n,k}] +  (\pm 1)^\ell {\sf N} [\overbrace{A_\ell N}{}_{n,k-1}] =  N_{n+1,k} \label{ZZZ}
 \end{align}
 Then:
\begin{align*}
&(\pm 1)^\ell A_\ell N_{n,0} = (\pm 1)^\ell {\sf N}[A_\ell N_{n,0}] + (\pm )^\ell {\sf N}[\overbrace{A_\ell N}{}_{n,0}]\\
&(\pm 1)^\ell A_\ell N_{n,1} = (\pm 1)^\ell {\sf N}[A_\ell N_{n,1}] + (\pm )^\ell {\sf N}[\overbrace{A_\ell N}{}_{n,1}]\\
&\ldots
 \end{align*}
 It is $ (\pm 1)^\ell {\sf N}[A_\ell N_{n,0}]={\sf N}[A_0...A_\ell... A_n] = N_{n+1,0}$. The other terms combine with \eqref{ZZZ}, and the  
sum proves the theorem.
 \end{proof} 
\end{thrm}
 As an interesting application, consider an $n$-particle Green function
\begin{align}
&i^n G(x_1\ldots x_n,\,y_1\ldots y_n)\\
&= \langle gs| {\sf T}\psi(x_1)\ldots \psi(x_n)\psi^\dagger (y_n)\ldots \psi^\dagger (y_1)|gs\rangle
\nonumber 
\end{align}
where $x$ denotes a complete set of quantum numbers and time, and the Heisenberg
evolution is given by the Hamiltonian whose ground state is $|gs\rangle $.
For independent particles Wick's teorem applies.
The average is evaluated as a sum of total T-contractions of field operators,
i.e. propagators (we now exclude anomalous propagators):
\begin{align}
&G^0(x_1\ldots x_n,\,y_1\ldots y_n) \nonumber\\
&=\sum_P (\pm 1)^P G^0(x_1,y_{i_1})\ldots G^0(x_n, y_{i_n})
\end{align}
where $P$ is the permutation $P(1\ldots n)=(i_1\ldots i_n)$. The sum
corresponds to the evaluation of the permanent (bosons) or determinant 
(fermions) of the matrix $ G^0(x_i,x_j)$, $ i,j=1,\ldots n$.

A free theory is a many-particle theory where, in some basis, $n$-particle Green functions are determined 
solely by one-particle Green functions.

\subsection*{Acknowledgement} I thank Dr. Daniel Ariad for helpful suggestions that clarified the original text.

\vfill
\end{document}